\documentclass[twocolumn,showpacs,amsmath,prl]{revtex4-1}

\usepackage{graphicx}
\usepackage{subfigure}
\usepackage{epsfig}
\usepackage{dcolumn}
\usepackage{bm}
\usepackage{color}
\usepackage{xcolor}

\usepackage{amsmath, amssymb, dsfont}
\usepackage{multirow}

\def \bea{\begin{eqnarray}}
\def \eea{\end{eqnarray}}

\def \ba{\begin{array}{cccc}}
\def \ea{\end{array}}
\def \nn{\nonumber}

\begin{document}

\title{Ideal Weyl semimetals in the chalcopyrites CuTlSe$_2$, AgTlTe$_2$, AuTlTe$_2$ and ZnPbAs$_2$}

\author{Jiawei Ruan$^{1\dagger}$, Shao-Kai Jian$^{2\dagger}$, Dongqin Zhang$^{1}$, Hong Yao$^{2,3,\ast}$, Haijun Zhang$^{1,}$}
\email{yaohong@tsinghua.edu.cn; zhanghj@nju.edu.cn. $^\dag$Equal contributions.}
\author{Shou-Cheng Zhang$^{4}$, Dingyu Xing$^1$}

\affiliation{
$^1$ National Laboratory of Solid State Microstructures, School of Physics, and Collaborative Innovation Center of Advanced Microstructures, Nanjing University, Nanjing 210093, China\\
$^2$ Institute for Advanced Study, Tsinghua University, Beijing 100084, China\\
$^3$ Collaborative Innovation Center of Quantum Matter, Beijing 100084, China \\
$^4$ Department of Physics, McCullough Building, Stanford University, Stanford, CA 94305-4045, USA
}


\begin{abstract}
Weyl semimetals are new states of matter which feature novel Fermi arcs and exotic transport phenomena. Based on first-principles calculations, we report that the chalcopyrites CuTlSe$_2$, AgTlTe$_2$, AuTlTe$_2$ and ZnPbAs$_2$ are ideal Weyl semimetals, having largely separated Weyl points ($\sim$ 0.05$\AA^{-1}$) and uncovered Fermi arcs that are amenable to experimental detections.  We also construct a minimal effective model to capture the low-energy physics of this class of Weyl semimetals.  Our discovery  is a major step toward a perfect playground of intriguing Weyl semimetals and potential applications for low-power and high-speed electronics.
\end{abstract}

\date{\today}



\maketitle

Weyl fermions, originally introduced as massless chiral fermions, are described by the Weyl equation\cite{weyl1929}. Even though a number of elementary particles were considered as candidates of Weyl fermions, conclusive evidences of Weyl fermions as elementary particles are still lacking. Weyl fermions were also proposed as emergent low-energy quasiparticles in condensed matter systems breaking either time-reversal or spatial-inversion symmetry \cite{nielsen1983,wan2011,Xu2011a,burkov2011,zyuzin2012,yang2011, halasz2012,hosur2013, zhang2014a, liu2014, weng2015, huang2015, soluyanov2015,hirayama2015, Ruan2016a}. One hallmark of Weyl semimetals is the existence of Fermi arcs in surface states\cite{wan2011}. So far the only experimentally known Weyl semimetals are the TaAs-class compounds, in which two sets of inequivalent Weyl points away from the Fermi level and complex Fermi surfaces were found by ARPES experiments\cite{xu2015, lv2015,yang2015}. Experimental evidences of negative magnetoresistance\cite{nielsen1983,son2013} induced by the chiral anomaly were also reported\cite{huang2015b,yang2015a,Shekhar2015,ong2015,li2015,li2015a}. However, definite signatures of the chiral anomaly in the quantum limit, such as the linear-$B$ negative magnetoresistance\cite{nielsen1983,son2013,zhang2015b} and the emergent supersymmetry\cite{jian2015}, and others\cite{wei2012,liu2013,ashby2013,sun2015} haven't been experimentally observed in known Weyl semimetals, which is partly due to the facts that the Weyl points are not all at the Fermi level and that there are coexisting trivial Fermi pockets. Therefore, it is urgent to discover ideal Weyl semimetals with {\it only} symmetry-related Weyl points at the Fermi level.

\begin{figure}[t]
\includegraphics[clip,angle=0,width=8cm]{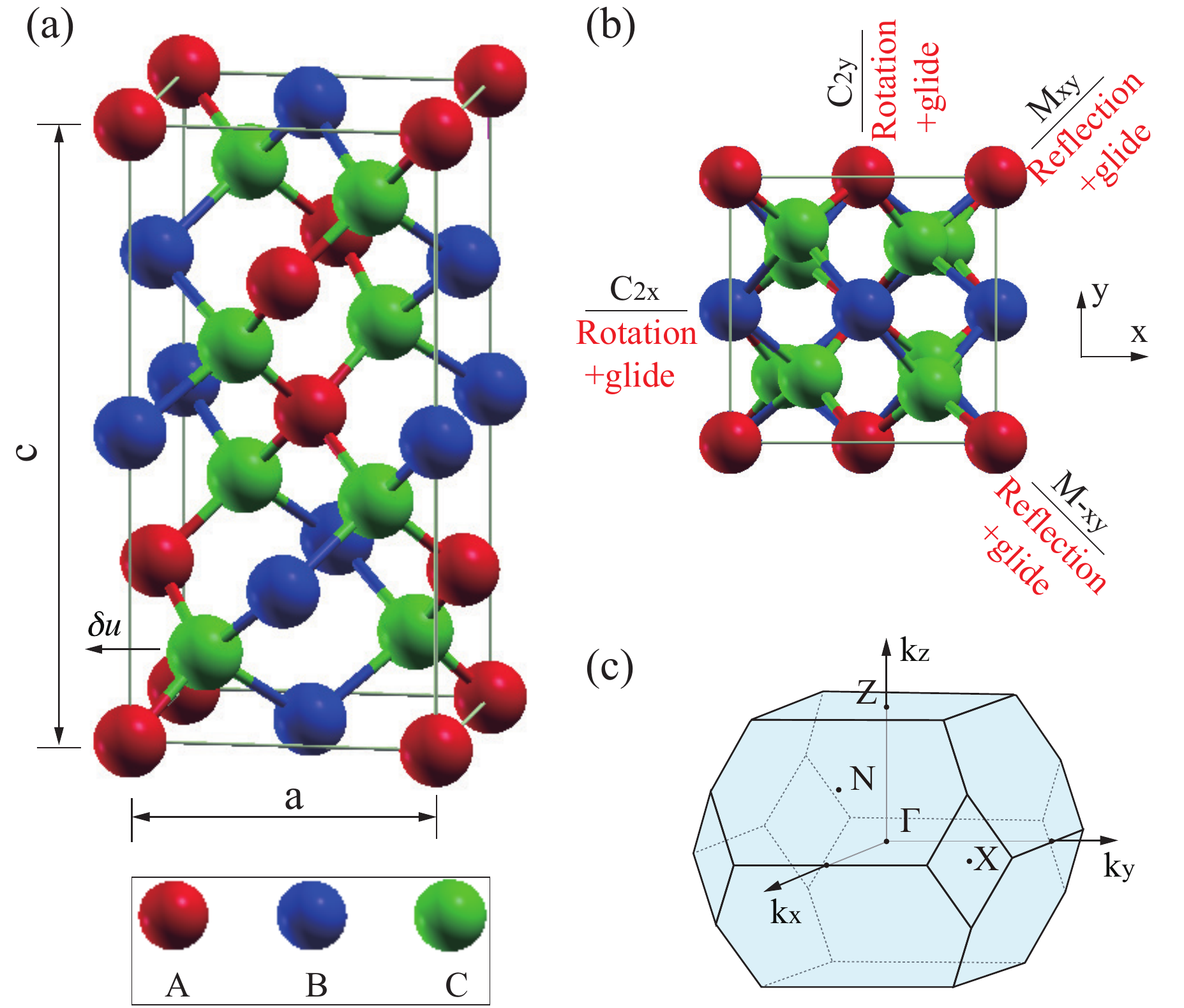}
\caption{Crystal structure and  Brillouin zone (BZ). (a) The crystal structure of chalcopyrite compounds ABC$_2$, including CuTlSe$_2$, CuTlTe$_2$, AgTlTe$_2$, AuTlTe$_2$, ZnPbAs$_2$ and ZnPbSb$_2$. The deviation of C atom (green) away from the center of tetrahedron formed by A and B atoms is denoted by $\delta u$.  (b) The top view of the chalcopyrite structure. The two-fold rotation symmetries ($C_{2x}$ and $C_{2y}$) and the two mirror symmetries ($M_{xy}$ and $M_{-xy}$) are marked.  (c) The BZ of chalcopyrite compounds. }
\end{figure}

\begin{figure}[t]
	\includegraphics[clip,angle=0,width=8.5cm]{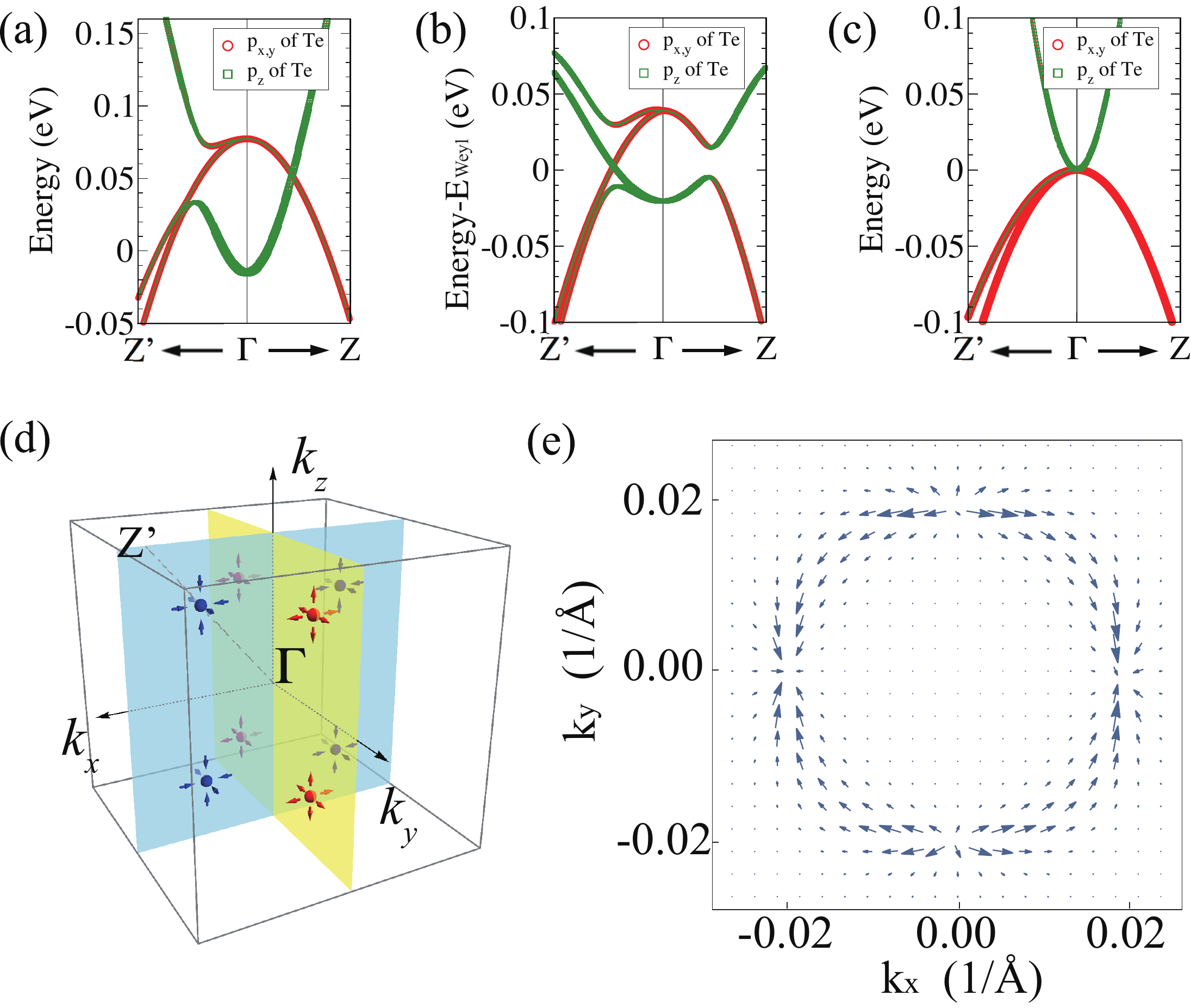}
	\caption{Band structure and Weyl points. (a) The bands of CuTlTe$_2$ without turning on  the spin-orbit coupling (SOC) effect.   The $p_z$ bands are below $p_{x,y}$ bands at the $\Gamma$ point. (b) The bands of CuTlTe$_2$ with turning on the SOC effect. An energy gap opens along the line of $\Gamma$-$Z$, and one Weyl point is seen between $\Gamma$ and $Z^{\prime}$. E$_{\textrm{Weyl}}$  denotes the energy level of Weyl points. (c) The bands of Hg$_2$Te$_2$ without turning on the SOC effect at the $\Gamma$ point are three-fold degenerate protected by the cubic symmetry.  (d) The schematic eight symmetry-related Weyl points in the BZ. (e) Berry curvature in a $k_z$ plane including four Weyl points. The Weyl points at $(\pm k^\ast_x,0,\pm k^\ast_z)$ have the `$-1$' chirality and at $(0,\pm k^\ast_y,\pm k^\ast_z)$ have `$+1$' chirality. }
\end{figure}

In this work, we focus on a large family of ternary chalcopyrites  ABC$_2$ at stoichiometry, which were of great interest because of potential applications including the thermoelectric effect, non-linear optics and solar cells\cite{shay1975,wei1995}. Recently, some ternary chalcopyrites were predicted to be topological insulators\cite{Feng2011}. Here, our first-principles calculations find that the chalcopyrite compounds CuTlSe$_2$, AgTlTe$_2$, AuTlTe$_2$ and ZnPbAs$_2$ are a class of ideal Weyl semimetals having eight symmetry-related Weyl points exactly at the Fermi level, but without any fine tuning. CuTlTe$_2$ and ZnPbSb$_2$  are also Weyl semimetals having eight symmetry-related Weyl points in energy gaps but  have a few coexisting trivial bands around the $X$ point; consequently they are not ideal Weyl semimetals at stoichiometry but can be tuned to be ideal Weyl semimetals by gating or doping.
The ideal Weyl semimetals predicted in the chalcopyrites are analogous to those  in compressively strained HgTe and half-Heusler compounds\cite{Ruan2016a}, but have one important advantage: external strain is no longer needed to realize ideal Weyl semimetal phases in these chalcopyrites. The surface Fermi arcs of these Weyl semimetals on both the (001) and (010) surfaces are 
uncovered 
and are amenable to ARPES detections. Such ideal Weyl semimetals provide a promising arena to observe the weak-field linear-$B$ negative magneto-resistance\cite{nielsen1983,zhang2015a}, a signature of the chiral anomaly in the quantum limit, which is still in debate so far. \\

\noindent{\bf Electronic structures}\\
\noindent The chalcopyrite compounds share a body-centered-tetragonal (bct) crystal structure with the space group D$_{2d}^{12}$ ($I\bar{4}2d$) which could be obtained by doubling the zinc-blende structure along the $z$ direction, such that its lattice constant $c$ is about twice of $a$, as shown in Fig. 1a. It has two twofold rotation symmetries $C_{2x}$ and $C_{2y}$ and two mirror symmetries $M_{xy}$ and $M_{-xy}$ combining with a proper glide (Fig.~1b). Generally, in the structure of chalcopyrites ABC$_2$, two A and two B atoms form a tetrahedron surrounding one C atom.  The tetrahedron is slightly tetragonally distorted, characterized by the ratio of the lattice constants $\eta=c/2a$, and the C atoms is off the center of the tetrahedron, characterized by $\delta u =(R_{AC}^2-R_{BC}^2)/a^2$ where $R_{AC}$ and $R_{BC}$ denote the distance between C and its nearest A and B atoms respectively.  Heuristically, the zinc-blende HgTe can be regarded as a special ``chalcopyrite'' ABC$_2$ with A=Hg, B=Hg, C=Te, $\eta=1$, and $\delta u =0$.  Interestingly, the chalcopyrite compounds ABC$_2$, where the cubic symmetry is broken due to the tetragonal distortion ($\eta \neq 1$) and the internal displacement ($\delta u \neq 0$),  can be effectively considered as a strained HgTe.

The I-III-VI$_2$ and II-IV-V$_2$ chalcopyrite compounds have sixteen valence electrons per unit cell, naively indicating an insulating ground state, which is true for CuInS$_2$ and ZnGeAs$_2$. However, an $sp$-type band inversion\cite{zhang2012a}, like in HgTe, exists in many I-III-VI$_2$ and II-IV-V$_2$ chalcopyrite compounds, which can give rise to topologically nontrivial state\cite{Feng2011}. Notably, as the chalcopyrite structure is effectively similar to a strained zinc-blende structure, those chalcopyrite compounds with the nontrivial band inversion are expected to realize topological insulators or ideal Weyl semimetals, depending on the type of the effective strain\cite{Ruan2016a}. Indeed, our first-principles calculations show that CuTlSe$_2$, AgTlTe$_2$, AuTlTe$_2$ and ZnPbAs$_2$ are ideal Weyl semimetals with eight symmetry-related Weyl points at the Fermi level, and CuTlTe$_2$ and ZnPbSb$_2$ are also Weyl semimetals with eight symmetry-related Weyl points but coexisting with trivial Fermi pockets. Four pairs of Weyl points are pinned either in the $k_x=0$ or $k_y=0$ plane because of crystal symmetries, schematically shown in Fig.~2d.

\begin{figure}[t]
\includegraphics[clip,angle=0,width=8cm]{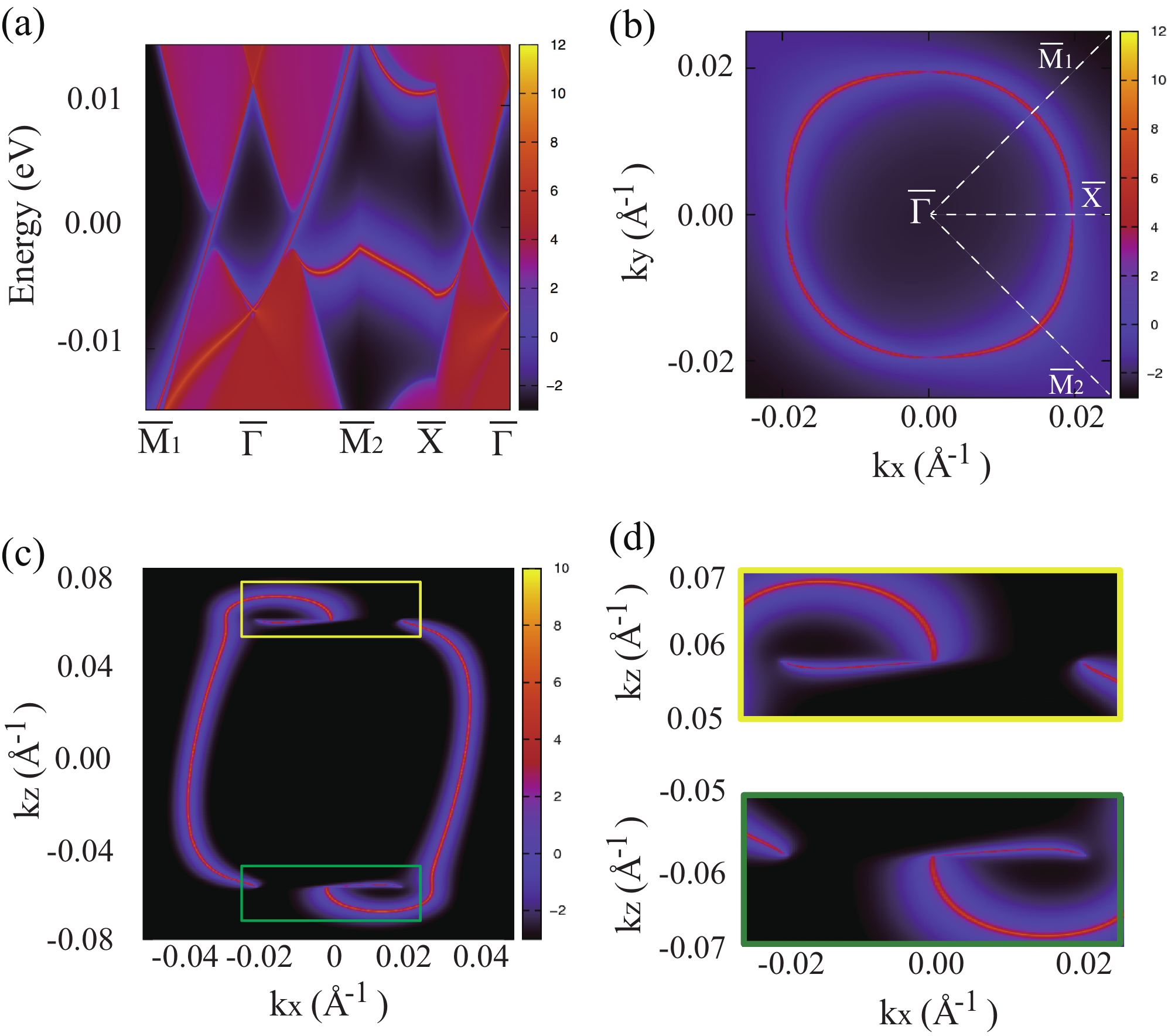}
\caption{ Surface states and Fermi arcs. (a) The local density of states (LDOS) for CuTlTe$_2$ projected onto the (001) surface. The warmer colors represent higher LDOS. The red/blue regions indicate bulk bands/energy gaps, and the single red lines indicate the surface states. The bulk band crossing ($\bar{k_x^\ast}$, 0), projected from the two Weyl points ($k_x^\ast$, 0, $\pm k_z^\ast$), can be seen in the $\bar{\Gamma}$-$\bar{X}$  line. The velocities of surface states along $\bar{\Gamma}$-$\bar{M}_1$ and  $\bar{\Gamma}$-$\bar{M}_2$ have opposite sign. (b) A closed Fermi surface consists of four Fermi arcs in the (001) surface. The high-symmetry lines  in (a) are marked.  (c) The open Fermi arcs on the (010) surface. (d) The zoomed-in Fermi arcs of (c).}
\end{figure}

Without loss of generality, we take CuTlTe$_2$ as an example to show the unique feature of symmetry-protected Weyl points in topologically nontrivial chalcopyrite compounds I-III-VI$_2$ and II-IV-V$_2$. The band inversion between the $s$ bands of Cu and Tl and the $p$ bands of Te around $\Gamma$, which makes the $p$ bands of Te dominate the Fermi level, is shown in Fig.~2a and b.   The bands of CuTlTe$_2$ are very similar to those of HgTe (equivalently Hg$_2$Te$_2$), except some features due to the tetragonal distortion in the chalcopyrite structure ($\eta \neq 1$ and $\delta u \neq  0$). We can see that the top of valence bands of Hg$_2$Te$_2$, without turning on the SOC effect,  are three-fold degenerate, indicating the degeneracy of $p_x$, $p_y$ and $p_z$ orbitals at the $\Gamma$ point, shown in Fig.~2c. Contrastively, in CuTlTe$_2$, the top of valence bands at $\Gamma$ split into two-fold degenerate bands ($p_{x,y}$) and one single band ($p_{z}$), shown in Fig.~2a, which indicates an effective tensile uniaxial strain along the $c$ direction.  Therefore, CuTlTe$_2$ is expected to be an ideal Weyl semimetal, like the strained HgTe\cite{Ruan2016a}. The Weyl point is really seen in the bands of CuTlTe$_2$ (Fig.~2b).  There are totally eight Weyl points which are all related to each other by symmetries, schematically shown in Fig.~2d. They are confined in $k_{x}=0$ or $k_{y}=0$ planes by the $C_{2T}=C_2\cdot T$ symmetry, where $C_2$ denotes a two-fold rotation $C_{2x}$ or $C_{2y}$, and $T$ denotes the time-reversal symmetry.  The four Weyl points in the $k_{x}=0$ ($k_{y}=0$) plane have the same chirality, for example, `$+1$' (`$-1$') in Fig.~2d and e.\\

\noindent {\bf Surface states and Fermi arcs}\\
\noindent The existence of topologically protected Fermi arcs is one hallmark of Weyl semimetals. We calculate the surface states and Fermi arcs of CuTlTe$_2$ through the maximally localized Wannier functions on the basis of first-principles calculations. The local density of states (LDOS) on the (001) surface is shown in Fig.~3a. One touching point  ($\bar{k_x^\ast}$, 0) in the $\bar{X}$-$\bar{\Gamma}$, projected from the two Weyl points ($k_x^\ast$, 0, $\pm k_z^\ast$), has a monopole charge `$-$2'.  Similarly, the touching point ($-\bar{k_x^\ast}$, 0) also has a monopole charge `$-2$', while both (0,$\pm\bar{k_y^\ast}$) have a monopole charge `$+2$'.  Therefore, one cannot see open Fermi arcs on the (001) surface Fermi surface because of the monopole charges of $\pm 2$. We calculate the Fermi arcs on the (001) surface, shown in Fig.~3b, where a closed Fermi surface consisting of four individual Fermi arcs is presented. Most interestingly, the surface states have the velocities of opposite sign along $\bar{\Gamma}$-$\bar{M}_1$ and  $\bar{\Gamma}$-$\bar{M}_2$, shown in Fig.~3a, indicating that the closed Fermi surface consisting of four Fermi arcs is topological different from the surface state of topological insulators such as Bi$_2$Se$_3$\cite{zhang2009}, where two Fermi velocities along $\bar{\Gamma}$-$\bar{M}_1$ and $\bar{\Gamma}$-$\bar{M}_2$ have the same sign.

\begin{figure}[t]
\includegraphics[clip,angle=0,width=8.5cm]{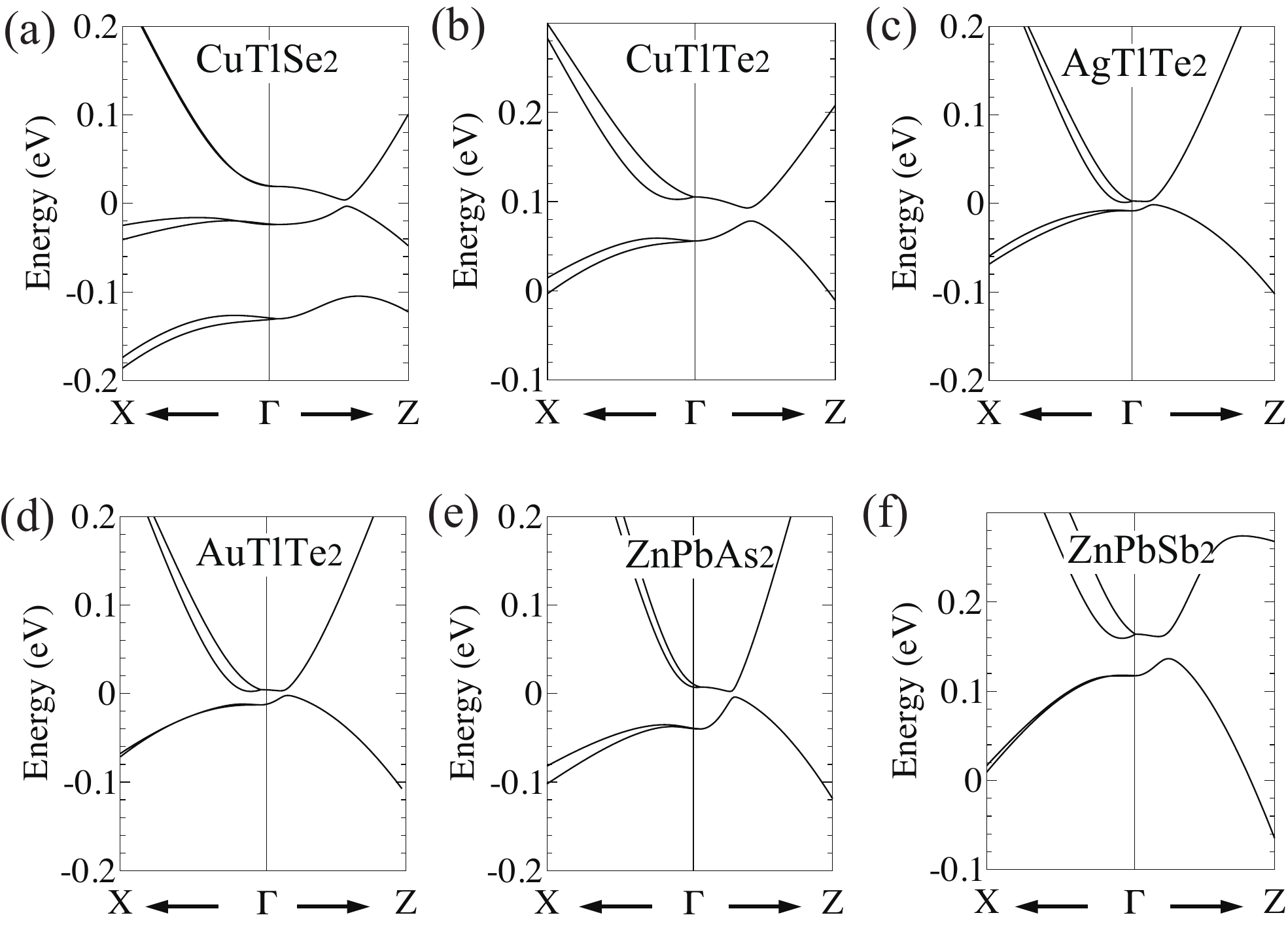}
\caption{ Band structures of Weyl semimetal candidates. (a-f) The band structures with turning on the SOC effect by the mBJ calculations for CuTlSe$_2$ (a), CuTlTe$_2$ (b), AgTlTe$_2$ (c), AuTlTe$_2$ (d), ZnPbAs$_2$ (e) and ZnPbSb$_2$ (f).   }
\end{figure}

On the (010) surface, there are six gapless points, including four points ($\pm\bar{k_x^\ast}$, $\pm\bar{k_z^\ast}$) with the monopole charge `$-1$', and two points (0,$\pm \bar{k_z^\ast}$) with the monopole charge `$+2$'. The open Fermi arcs emerge, as shown in Fig.~3c, and the zoomed-in parts of Fermi arcs are shown in Fig.~3d. The Fermi arcs are consistent with those in strained HgTe \cite{Ruan2016a}. However, it is worth noticing that the Weyl points here are well separated, for example, about 5\% of the reciprocal lattice constant between the two Weyl points ($\pm k_x^\ast$, 0, $k_z^\ast$), which is much  amenable for experimental detections.\\

\noindent {\bf Weyl semimetal candidates}\\
\noindent In Fig.~4, based on the modified Becke-Johnson (mBJ) calculations, the bulk band structures of six chalcopyrite compounds CuTlSe$_2$,  CuTlTe$_2$, AgTlTe$_2$, AuTlTe$_2$, ZnPbAs$_2$, and ZnPbSb$_2$, are shown, exhibiting similar band structures. Based on our calculations, they are all Weyl semimetals with eight symmetry-protected Weyl points, especially including four ideal Weyl semimetals CuTlSe$_2$, AgTlTe$_2$, AuTlTe$_2$, and ZnPbAs$_2$. For CuTlTe$_2$ and ZnPbSb$_2$, the bottom of the conduction bands stays around the $X$ point at the boundary of the BZ (seen in Fig.~S2c in the supplementary information(SI)) which is far from the $\Gamma$ point. Therefore, although there are trivial bulk bands coexisting with the Weyl points, the Weyl points are still well separated from the trivial bands in CuTlTe$_2$ and ZnPbSb$_2$.  For all these six Weyl semimetals, the positions of Weyl points ($\pm k_x^\ast$, 0, $\pm k_z^\ast$) and  (0, $\pm k_y^\ast$, $\pm k_z^\ast$) are calculated by both standard GGA and mBJ calculations, listed in Table S1 in SI. We can see that the larger difference in atomic radius between A and B atoms in chalcopyrites ABC$_2$ produces the larger $k_z^\ast$ for Weyl points, and the larger SOC effect of C atom results in larger $k_{x,y}^\ast$. Consequently, Weyl points of CuTlTe$_2$ and ZnPbSb$_2$ are the most separated among these materials.\\

\noindent {\bf Low-energy effective model}\\
\noindent The topological nature in these ideal Weyl semimetals is determined mainly by the low-energy physics near the Weyl points, which can be captured by a low-energy effective Hamiltonian. Up to
the quadratic order of momenta ${\bf k}$, the low-energy effective model can be explicitly written down under the restriction to the time-reversal symmetry and the lattice symmetry (D$_\text{2d}$), as shown in SI. Interestingly, we can reach an analytically-trackable low-energy effective model which can capture the topological features of these ideal Weyl semimetals:
\begin{eqnarray}
&&H({\bf k}) = \epsilon_0({\bf k}){\rm I}_{4\times 4}+ c_1 (k_y k_z \Gamma^1+ k_z k_x \Gamma^2)+c_2 k_x k_y \Gamma^3 \\
&&+c_4 (k_x^2-k_y^2)\Gamma^4\!+\!\big[c_3 (k_z^2-m^2)+c_5(k_x^2+k_y^2)\big] \Gamma^5 \!+\!v k_z \Gamma^{35},\nn
\end{eqnarray}
where $\epsilon_0({\bf k})=a_0+a_1 (k_x^2+k_y^2)+a_2 k_z^2$, $\Gamma^i$ are $4\!\times\!4$ matrices given in the SI, and $c_i$, $a_i$, $m$, $v$ are constants. The effective model can be understood by lowering the symmetry of Luttinger Hamiltonian. The $m^2 \Gamma^5$ and $k_z \Gamma^{35}$ terms break reflection and inversion symmetries, respectively, rendering a quadratic band-touching to split into eight Weyl points in the high-symmetry plane as stated before. The eigen-energy reads 
$E({\bf k})=\epsilon_0({\bf k})\pm \big[d_1^2({\bf k})+d_2^2({\bf k})+d_3^2({\bf k})\big]^\frac12$, where $d_i({\bf k})$ are given in the SI. There are eight Weyl points in this effective Hamiltonian whose locations are given in the SI for certain range of parameters.  
The projected Hamiltonian near one of the Weyl points is described by the Weyl equations, i.e.,
\bea
	H_{\text{Weyl}}({\bf k})= -c_1 m k_x \sigma^x + \frac{c_2 v}{c_1} k_y \sigma^y+ 2c_3 m k_z \sigma^z,
\eea
where the identity matrix part is neglected because it does not affect the topological properties. Since the low-energy physics is captured by the locations and the Fermi velocities of the Weyl points which can be determined uniquely by the the effective model, it correctly captures the low-energy physics in this class of Weyl semimetals. \\

\noindent {\bf Acknowledgement}\\
We appreciate G. Yao for technical supports and  Y. M. Pan for helping to prepare figures. H.Y. is supported in part by the National Thousand-Young-Talents Program and by the NSFC under Grant No. 11474175 at Tsinghua University. H.J.Z is supported by the Scientific Research Foundation of Nanjing University (020422631014) and the National Thousand-Young-Talents Program. S.C.Z. is supported by  the Department of Energy, Office of Basic Energy Sciences, Division of Materials Sciences and Engineering, under contract DE-AC02-76SF00515 and by FAME, one of six centers of STARnet, a Semiconductor Research Corporation program sponsored by MARCO and DARPA.

\bibliography{TI}

\begin{widetext}

\section{supplementary information}

\renewcommand{\theequation}{S\arabic{equation}}
\setcounter{equation}{0}
\renewcommand{\thefigure}{S\arabic{figure}}
\setcounter{figure}{0}
\renewcommand{\thetable}{S\arabic{table}}
\setcounter{table}{0}

\subsection{Appendix A: The methods of the first-principles calculations}

The {\it ab-initio} calculations are carried out in the framework of the Perdew-Burke-Ernzerhof-type generalized gradient approximation of the density functional theory through employing the BSTATE package\cite{fang2002} with the plane-wave pseudo-potential method. The kinetic energy cutoff is fixed to 340eV, and the {\bf k}-point mesh is taken as 16$\times$16$\times$16 for the bulk calculations. The spin-orbit coupling effect is self-consistently included. The modified Becke-Johnson (mBJ) calculations \cite{tran2009} are used to correct the underestimated band gaps. All the calculations are further confirmed by the Vienna Ab initio simulation package (VASP).\cite{kresse1996} The lattice constants and atoms are fully relaxed with the total energy cutoff of $1.E-7$ eV. Maximally localized Wannier functions\cite{marzari1997,souza2001} are employed to obtain the {\it ab initio} tight-binding model of semi-infinite systems with the (001) or (010) surface as the  boundary\cite{Zhang2009a} to exhibit topological surface states and Fermi arcs. An iterative method \cite{sancho1984,sancho1985} is used to obtain the surface Green's function of the semi-infinite system.

\subsection{Appendix B:  Band structures of CuTlSe$_2$, CuTlTe$_2$, AgTlTe$_2$, AuTlTe$_2$, ZnPbAs$_2$ and ZnPbSb$_2$}

Band structures of CuTlSe$_2$, CuTlTe$_2$, AgTlTe$_2$, AuTlTe$_2$, ZnPbAs$_2$ and ZnPbSb$_2$ along more high-symmetry lines are calculated for both without and with turning on the SOC effect, and shown in Fig.~S1 and S2 respectively. CuTlSe$_2$, AgTlTe$_2$, AuTlTe$_2$ and ZnPbAs$_2$ present the ideal Weyl semimetals pictures without trivial bulk bands crossing the Fermi level. Differently, the bottom of valence bands of CuTlTe$_2$ and ZnPbSb$_2$ drop down to cross the Fermi level at around X point and push the Weyl nodes above the Fermi level.

\begin{figure}[t]
\includegraphics[width=16cm]{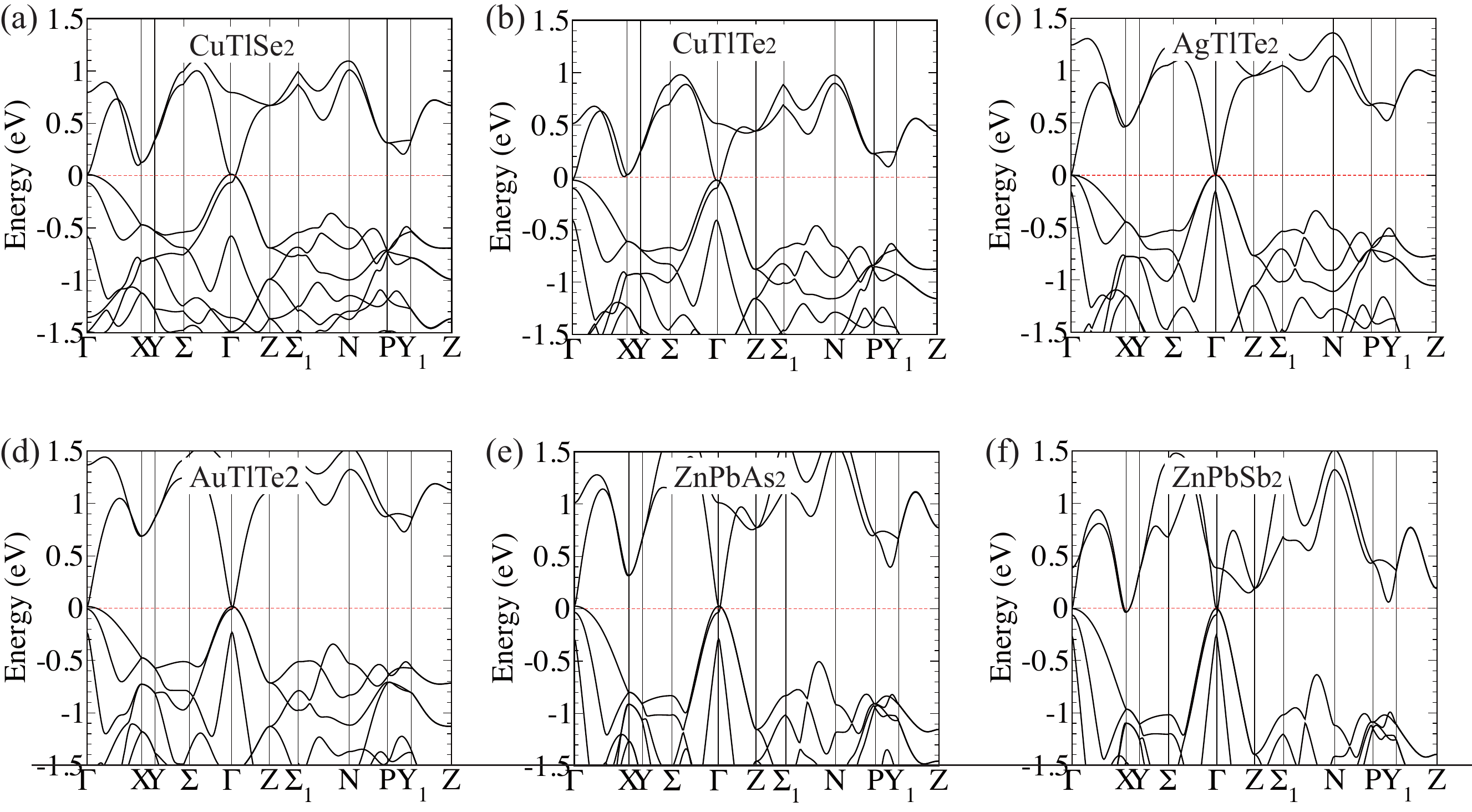}\label{fig-s1}\\
\caption{Band structures of  CuTlSe$_2$, CuTlTe$_2$, AgTlTe$_2$, AuTlTe$_2$, ZnPbAs$_2$ and ZnPbSb$_2$ along high symmetry lines $\Gamma$-X-Y-$\Sigma$-$\Gamma$-Z-$\Sigma_1$-N-P-Y$_1$-Z, without turning on the SOC effect.}
\end{figure}

\begin{figure}[t]
\includegraphics[width=16cm]{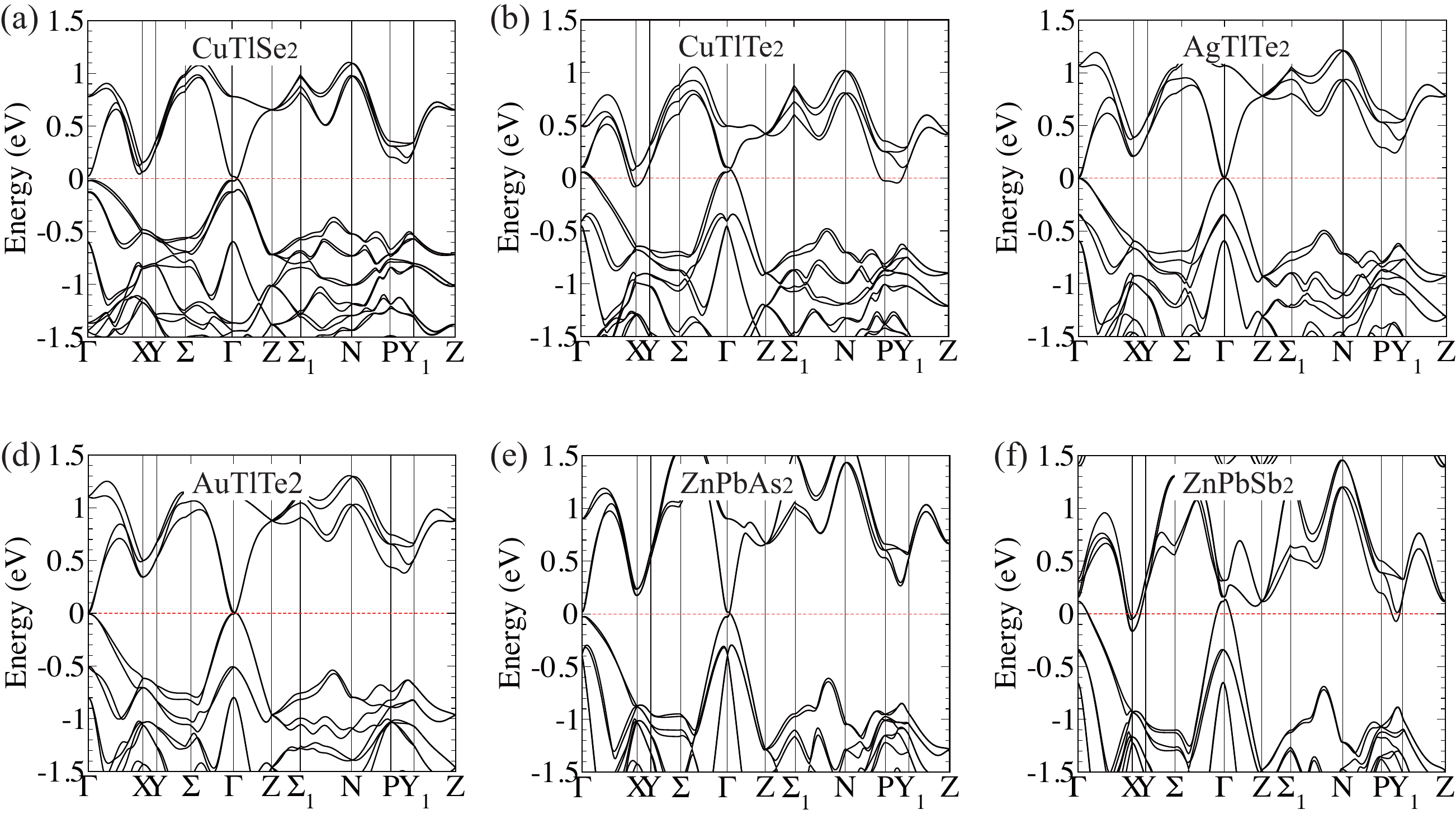}\label{fig-s2}\\
\caption{ Band structures of  CuTlSe$_2$, CuTlTe$_2$, AgTlTe$_2$, AuTlTe$_2$, ZnPbAs$_2$ and ZnPbSb$_2$ along high symmetry lines $\Gamma$-X-Y-$\Sigma$-$\Gamma$-Z-$\Sigma_1$-N-P-Y$_1$-Z, with turning on the SOC effect. }
\end{figure}

\subsection{Appendix C: The Weyl points in CuTlSe$_2$, CuTlTe$_2$, AgTlTe$_2$, AuTlTe$_2$, ZnPbAs$_2$ and ZnPbSb$_2$}

Based on both GGA and mBJ calculations, chalcopyrite compounds CuTlSe$_2$, CuTlTe$_2$, AgTlTe$_2$, AuTlTe$_2$, ZnPbAs$_2$ and ZnPbSb$_2$ are found to be Weyl semimetal candidates with four pairs of Weyl points in the BZ, including four ideal Weyl semimetal candidates (CuTlSe$_2$, AgTlTe$_2$, AuTlTe$_2$ and ZnPbAs$_2$). GGA calculations underestimate the band gap, which means the band inversion is overestimated, but the mBJ calculations could much correct the band energy, so the conclusion of the ideal Weyl semimetals, in Table S1, is based on mBJ calculations. For the others, such as, the locations of Weyl points and the separation between Weyl points, we give all results from both GGA and mBJ to compare.

\begin{table}[b]
\caption{The locations of the four pairs of Weyl points in CuTlSe$_2$, CuTlTe$_2$, AgTlTe$_2$, AuTlTe$_2$, ZnPbAs$_2$ and ZnPbSb$_2$ are ($\pm k_x^\ast$, 0, $\pm k_z^\ast$) and (0, $\pm k_y^\ast$, $\pm k_z^\ast$) in BZ with $k_x^\ast=k_y^\ast$ because of the mirror symmetry. The volume of `\% of 2$\pi$/a' presents $2k_x/(2\pi/a)\times100\%$, where $a$ denotes the in-plane lattice constant,  indicating how much separated the Weyl points are. `Ideal Weyl' means the ideal Weyl semimetals. `GGA' denotes the common generalized gradient approximation of density functional theory, and `mBJ' denotes the modified Becke-Johnson (mBJ) method.}

\begin{ruledtabular}
\begin{tabular}{ccccccccc}
\multirow{2}*{} & \multicolumn{3}{c}{GGA} & \multicolumn{4}{c}{mBJ} & \multirow{2}*{Ideal Weyl
} \\
\cline{2-4}\cline{5-8}
& $k_x^\ast=k_y^\ast(\textrm{\AA}^{-1})$ & $k_z^\ast(\textrm{\AA}^{-1})$ & $\%$ of $2\pi/a$ & $k_x^\ast=k_y^\ast(\textrm{\AA}^{-1})$ & $~~k_z^\ast(\textrm{\AA}^{-1})~~$ & $\%$ of $2\pi/a$ & E$_{\textrm{Weyl}}(\textrm{eV})$&  \\
\colrule
  $\text{CuTlSe}_2$ & 0.0061 &0.0695 & 1.2 & 0.0069 & 0.0589 & 1.4 &0.000 & Y \\
  $\text{CuTlTe}_2$ & 0.0202 &0.0576 & 4.0 & 0.0115 & 0.0483 & 2.2 & 0.085& N \\
  $\text{AgTlTe}_2$ & 0.0056 &0.0119 & 1.1 & 0.0049 & 0.0117 & 1.1 & 0.000& Y \\
  $\text{AuTlTe}_2$ & 0.0376 &0.0493 & 7.5 & 0.0058 & 0.0192 & 1.2 & 0.000& Y \\
  $\text{ZnPbAs}_2$ & 0.0054 &0.0416 & 1.1 & 0.0035 & 0.0268 & 0.7 & 0.000& Y \\
  $\text{ZnPbSb}_2$ & 0.0134 &0.0397 & 2.6 & 0.0133 & 0.0254 & 2.6 & 0.145& N \\

\end{tabular}
\end{ruledtabular}
\end{table}

\begin{table}[t]
\caption{The summary of parameters in low-energy effective model in Eq. \ref{kp_full} for CuTlTe$_2$.}

\begin{ruledtabular}
\begin{tabular}{cccc}

 parameters & values  & parameters & values\\
\colrule
$a_0(\textrm{eV})$ & 0.081 & $c_5(\textrm{eV})$ & 5.37 \\
$a_1(\textrm{eV}\textrm{\AA}^2)$ & 1.72 & $m(\textrm{\AA}^{-1})$ & 0.058 \\
$a_2(\textrm{eV}\textrm{\AA}^2)$ & -2.05 & $v(\textrm{eV}\textrm{\AA})$ & 0.187 \\
$c_1(\textrm{eV}\textrm{\AA}^2)$ & -9.27 & $\alpha_1(\textrm{eV}\textrm{\AA})$ & 0.022 \\
$c_2(\textrm{eV}\textrm{\AA}^2)$ & -7.09 & $\alpha_2(\textrm{eV}\textrm{\AA})$ & -0.031 \\
$c_3(\textrm{eV}\textrm{\AA}^2)$ & -8.75 & $\alpha_3(\textrm{eV}\textrm{\AA})$ & -0.083 \\
$c_4(\textrm{eV}\textrm{\AA}^2)$ & -9.21 &  &  \\

\end{tabular}
\end{ruledtabular}
\end{table}

\subsection{Appendix D: The effective Hamiltonian}

\begin{figure}[t]
\includegraphics[width=10cm]{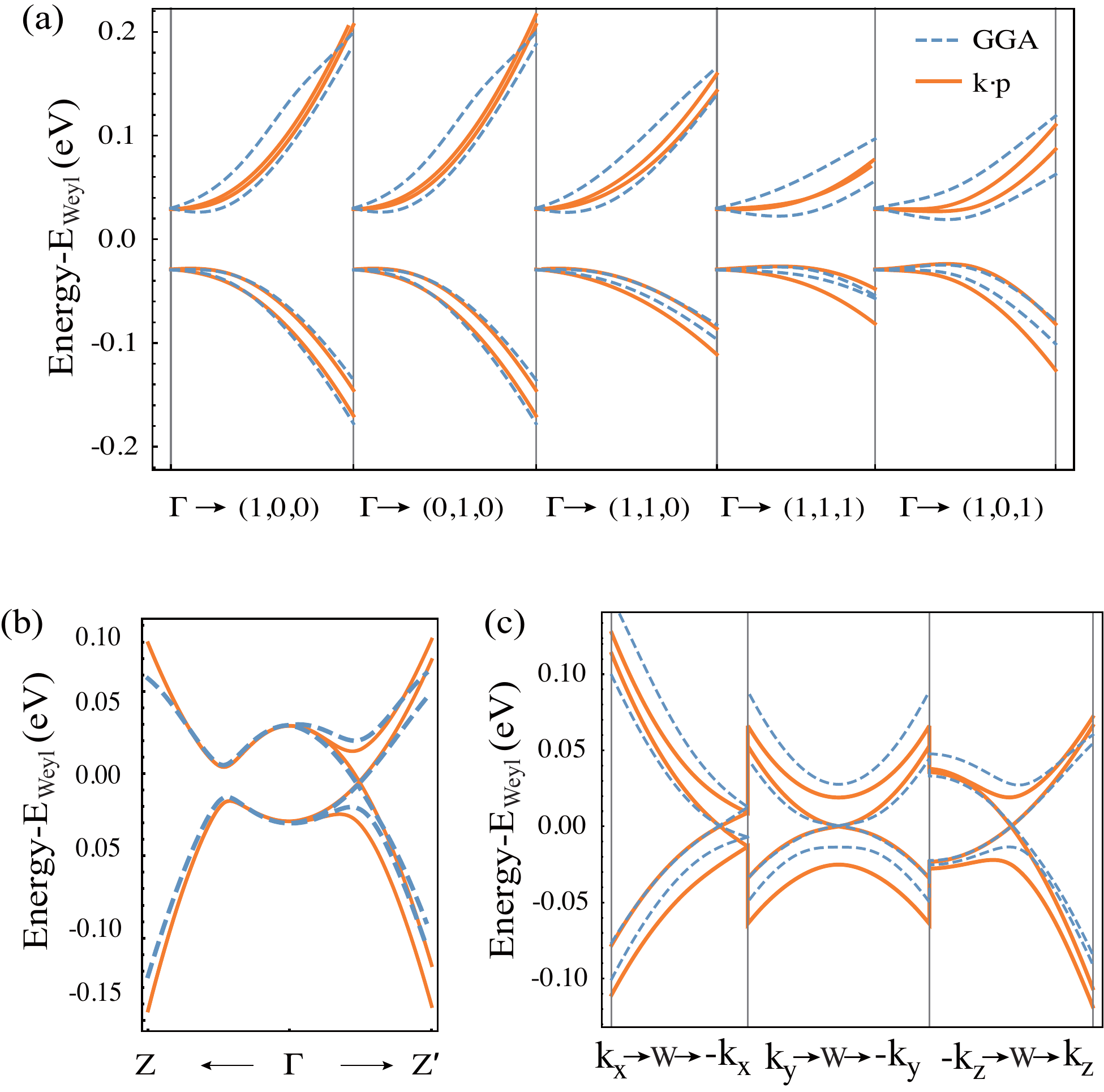}\label{fig-s3}\\
\caption{The bands obtained from the effective Hamiltonian in Eq. \ref{kp_full} (orange solid line) compared with those from first-principles calculations of CuTlTe$_2$ (blue dashed line). (a)The bands along different directions. (b)The bands along $\Gamma$-$Z$ and $\Gamma$-$Z'$. One Weyl point is seen in $\Gamma$-$Z'$. (c)The band dispersion along $-k_x$,$k_y$ and $k_z$ directions from the Weyl point at ($k_x^\ast,0,k_z^\ast)$, where `W' denotes the Weyl point. }
\end{figure}

The low-energy physics in the chalcopyrite is dominated by $p$-state electrons of Te atoms. In the presence of spin orbit coupling, those states with the total angular momentum $J=3/2$ contribute the most degrees of freedom near the Fermi energy. Thus we construct the effective Hamiltonian from $J=3/2$ multiplets. As stated in the main text, the effective model near the $\Gamma$ point is dictated by the lattice symmetry ($D_{\text{2d}}$ ), and the time reversal symmetry. All 4$\times$4 Hermitian matrices can be represented by sixteen $\Gamma$ matrices \cite{murakami2004}, forming five irreducible representations of D$_{2d}$ group:
\bea
		A_1&:& 1,\Gamma^5, \\
		A_2&:& \Gamma^{12},\Gamma^{34}, \\
		B_1&:& \Gamma^4, \Gamma^{45},  \\
		B_2&:& \Gamma^3,\Gamma^{35}, \\
		E&:& (\Gamma^2,\Gamma^1);(\Gamma^{15},\Gamma^{25});  (\Gamma^{23},\Gamma^{13}) ;( \Gamma^{14},-\Gamma^{24}).
\eea
Also, the momenta up to quadratic level could furniture four irreducible representations of D$_{2d}$ group:
\bea
		A_1&:& k_x^2+k_y^2,k_z^2, \\
		B_1&:& k_x^2-k_y^2, \\
		B_2&:& k_z,k_x k_y, \\
		E    &:& k_x,k_y; k_x k_z, k_y k_z.
\eea
Thus, the most general effective Hamiltonian up to quadratic level near the $\Gamma$ point is given by:
\bea
	H({\bf k}) &=&\epsilon_0({\bf k}){\rm I}_{4\times 4}+ c_1 (k_y k_z \Gamma^1+ k_z k_x \Gamma^2) +c_2 k_x k_y \Gamma^3+ [c_3 (k_z^2-m^2)+c_5(k_x^2+k_y^2)] \Gamma^5+c_4( k_x^2- k_y^2) \Gamma^4+ v k_z \Gamma^{35} \nn \\
&&+\alpha_1(k_x \Gamma^{15}+k_y \Gamma^{25})+\alpha_2(k_x\Gamma^{23}+k_y\Gamma^{13})
+\alpha_3(k_x\Gamma^{14}-k_y\Gamma^{24}). \label{kp_full}
\eea
where $\epsilon_0({\bf k})=a_0+a_1(k_x^2+k_y^2)+a_2 k_z^2$. The various constants $a_i$, $c_i$,$v$,$m$ and $\alpha_i$ describe the specific band properties. 
For a reference, we fit these constants to the results of first-principle calculations of CuTlTe$_2$ with the  normal GGA , and the summary is listed in Table S2. The bands from this effective model are calculated and compared with those from first-principles calculations in Fig.~S1. If we keep the linear term of $vk_z \Gamma^{35}$ and neglect the other linear terms in Eq. S10, an analytical energy dispersion can be obtained,
\begin{eqnarray}
E({\bf k})&=&\epsilon_0({\bf k})\pm \sqrt{d^2_1({\bf k})+d_2^2({\bf k})+d^2_3({\bf k})}.
\end{eqnarray}
where $d_1({\bf k})=c_2 k_x k_y$, $d_2({\bf k})= |v k_z|\pm \sqrt{c_4^2(k_x^2-k_y^2)^2+c_1^2(k_x^2+k_y^2)k_z^2}$, and $d_3({\bf k})= c_5(k_x^2+k_y^2)+c_3 (k_z^2-m^2)$.
The band-crossing points are given by the following conditions: $d_i({\bf k})=0$, $i=1,2,3$.
Apparently, there are eight solutions,$(\pm k_x^*,0,\pm k_z^*)$ and $(0,\pm k_y^*,\pm k_z^*)$ for $c_3 c_5<0$ (as expected from perturbing the Luttinger Hamiltonian), $v\ne 0$, and $ m \ne 0$, where $k_x^{*2}=k_y^{*2}=\bigg[-(c_1^2m^2+\frac{v^2c_5}{c_3})+\sqrt{(c_1^2m^2+\frac{v^2c_5}{c_3})^2+4m^2 v^2(c_4^2-\frac{c_1^2 c_5}{c_3})}\bigg]/(2c_4^2-2\frac{c_1^2 c_5}{c_3})$ and $k_z^{*2}= m^2-\frac{c_5 }{c_3}k_x^{*2}$. 
To address the topological properties of these band-crossing points, we project the Hamiltonian into one of those points. The downfolded two-bands model reads,
\bea
  H_{\text{Weyl}}= -c_1 m k_x \sigma^x + \frac{c_2 v}{c_1} k_y \sigma^y+ 2c_3 m k_z \sigma^z,
\eea
where the identity part is neglected since it does not affect the topological properties. Clearly, above Hamiltonian describes a Weyl fermions with the chirality given by $\chi= \text{sign}(- v c_2 c_3)$. In the ideal Weyl semimetal phase, the low energy physics is determined by the locations and the Fermi velocities of Weyl points. Therefore, the effective model in Eq. \ref{kp_full} can capture the topological properties of this class of ideal Weyl semimetals.

\end{widetext}

\end{document}